\documentclass[runningheads]{llncs}
\usepackage[T1]{fontenc}
\usepackage{amsmath}
\usepackage{amsfonts}
\usepackage{booktabs}
\usepackage{multirow}
\usepackage{multicol}
\usepackage{array}
\usepackage{subfig}
\usepackage{graphicx}
\usepackage[numbers,sort&compress]{natbib}
\usepackage{xspace}
\usepackage{hyperref}

\usepackage{color}

\begin{document}
\title{A Hybrid Approach for EMF Code Generation: Code Templates Meet Large Language Models}
\titlerunning{A Hybrid Approach for EMF Code Generation}

\author{Xiao He\inst{1} \and
Ru Chen\inst{1} \and
Zeqing Zhang\inst{1} \and
Yanling Wang\inst{2} \and
Qiuyan Dong\inst{2}
}
\authorrunning{X. He et al.}
%
\institute{School of Computer and Communication Engineering,\\ University of Science and Technology Beijing, Beijing 100083, China \and
Beijing Institute of mechanical and Electrical Engineering, Beijing, China\\
\email{hexiao@ustb.edu.cn}}

\newcommand{\iecoregen}{\textsc{iEcoreGen}\xspace}

\newcommand{\passk}{\textit{pass@k}\xspace}
\newcommand{\compk}{\textit{compilation@k}\xspace}
\newcommand{\passrate}{\textit{pass\_rate}\xspace}
\newcommand{\valdown}[1]{{\scriptsize{(\textcolor{blue}{$\downarrow$#1})}}\xspace}
\newcommand{\valup}[1]{{\scriptsize{(\textcolor{red}{$\uparrow$#1})}}\xspace}
\newcommand{\valpar}{{\scriptsize{($-$)}}\xspace}
\newcommand{\slfrac}[2]{\left.#1\middle/#2\right.}

\maketitle              
\begin{abstract}

Template-based and LLM-based code generation are both key enablers of automated software development.
The former provides correctness guarantees but are rigid for complex requirements, whereas LLMs offer high flexibility at the risk of producing faulty code.
This paper proposes \iecoregen, a hybrid approach that integrates Eclipse Modeling Framework (EMF) and LLMs. 
In EMF, an Ecore model defines a system structure and acts as a blueprint for code-generation. 
\iecoregen decomposes requirements to derive operation specifications, uses EMF's template-based generator to produce initial Java code, and serializes specifications into docstrings. 
LLMs are then invoked to complete and fix unimplemented methods.
We assessed \iecoregen on twenty code-generation tasks across five LLMs.
It surpasses LLM-only baselines on \passk and performs on par with them on \compk. 
An ablation study clarified the contribution of each component of \iecoregen. 
Overall, the findings indicate that LLM-enhanced model-driven development is a promising path toward more efficient software automation.


\end{abstract}

\section{Introduction}


Model-driven engineering (MDE) \cite{mda} has been a pivotal approach for automating software development \cite{automation_mde} over the past two decades.
When using this model-centric methodology, developers create software models that define the system's structure and behavior at a high level of abstraction. 
System implementation can be automatically generated by applying model transformation and code generation \cite{mdecg,cgbymt}.
MDE has been successfully employed in many domains \cite{mdeapplication,mde4web,mde4emb,mde4rob} and serves as the technical foundation of low-code development \cite{mde4lcd}.

\begin{figure}[!tb]
    \centering
    \includegraphics[width=1\linewidth]{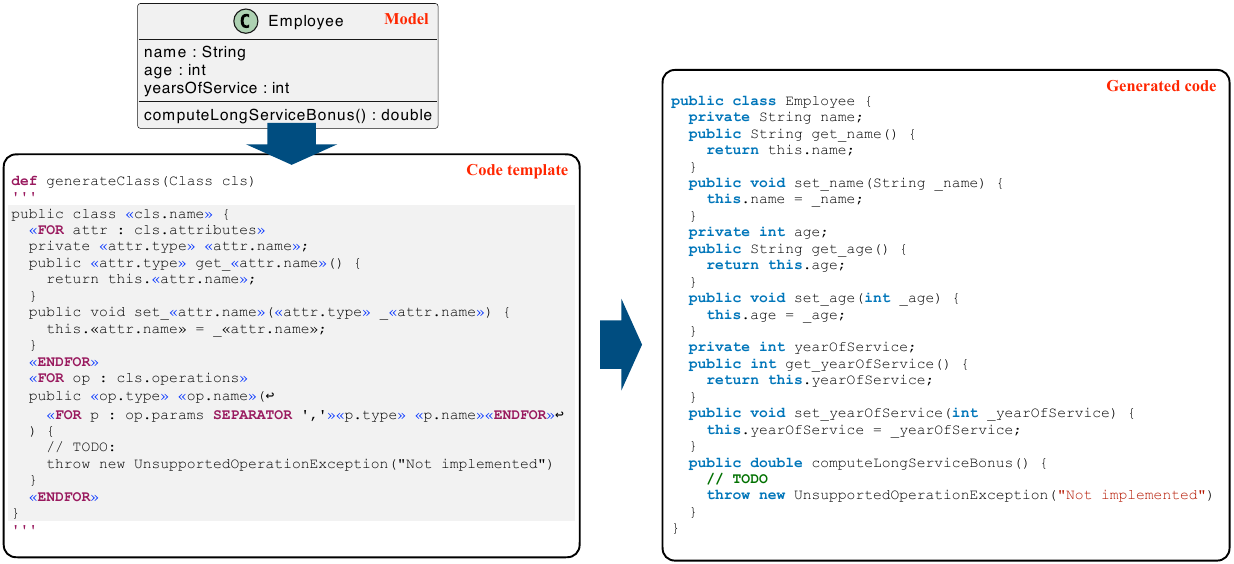}
    \caption{Code generation in MDE}
    \label{fig:codegen-mde}
\end{figure}

In Model-Driven Engineering (MDE), code generation is typically accomplished through code templates---programs that algorithmically transform model elements into code by filling in a predefined structure. 
To illustrate, Figure \ref{fig:codegen-mde} depicts the generation of a Java class from a UML \texttt{Employee} class. 
The Xtend template \texttt{generateClass} takes the UML class as input and outputs the corresponding Java code. 
While this approach, known as Template-Based Code Generation (TBCG), relies on meticulously crafted and verified rules to guarantee correct output for covered scenarios, this very strength is also its primary weakness. 
TBCG fails when system requirements \textit{fall outside the range} of its predefined templates. 
Our example demonstrates this limitation: while the template correctly generates getters and setters, it cannot implement the \texttt{computeLongServiceBonus()} operation's logic. 
Consequently, it produces an invalid method body that throws an \texttt{UnsupportedOperationException}.

Large language models (LLMs) are revolutionizing software engineering \cite{llm4se} by demonstrating remarkable capabilities in addressing various tasks, including code generation from natural language (NL) requirements \cite{llmcg-survey1,llmcg-survey2}. 
Being trained on billions of parameters from massive text corpora \cite{llmsurvey}, LLMs excel at understanding and generating code. 
A striking demonstration was OpenAI's GPT-5 outperforming all human contestants at ICPC 2025 by solving every problem\footnote{\url{https://venturebeat.com/ai/google-and-openais-coding-wins-at-university-competition-show-enterprise-ai}}. 
Yet LLMs still face major difficulties in complex code-generation tasks. They handle standalone statements and functions well but falter on multi-class or project-scale code \cite{javabench}, which demands reasoning across many modules within a large context.
This limitation exacerbates the model's tendency to hallucinate \cite{Zhang2025}, frequently resulting in non-functional or incorrect code.

Template-based and LLM-based code generation are not competing paradigms but complementary ones. 
This synergy immediately suggests a promising path forward: \textit{a hybrid approach that merges the reliability of templates with the adaptability of LLMs}. 
This paper introduces \iecoregen, a novel method that realizes this vision within MDE based on the Eclipse Modeling Framework \cite{emf} (EMF). 
\iecoregen combines template-based generation and LLMs to leverage their combined strengths. 
It accepts an Ecore model, a simplified class diagram supported by EMF, and an NL system requirement. 
The process starts by identifying necessary operations from the Ecore model and breaking down the system requirements into detailed, method-level NL specifications. 
\iecoregen then uses EMF's template-based code generation to convert the Ecore model into Java, embedding method-level specifications as Java docstrings. 
LLMs are then instructed to fill in method bodies based on these docstrings, conforming to the EMF-generated code structure. 
If compilation errors arise, \iecoregen continually directs LLMs to correct the code until it compiles successfully.

To evaluate, we performed an experiment using a benchmark of twenty problems, each featuring an NL requirement, a class diagram, and an NL test specification. 
We compared \iecoregen against three baselines that solely use LLMs. 
The experiment was conducted on five mainstream open-source LLMs, including Deepseek, Qwen3, GPT-OSS, Mistral, and Llama-4.
Preliminary results indicate that \iecoregen generally generates code with higher accuracy and outperforms the baseline approaches in terms of \passk, which implies an improvement in functional accuracy, without introducing extra compilation errors. 

This paper's main contributions include: 
\begin{itemize} 
    \item \textbf{\iecoregen}, an innovative hybrid approach combining TBCG and LLMs. 
    \item \textbf{A prototype} using EMF and Ecore that benefits the MDE community. 
    \item \textbf{An empirical study} on five open-source LLMs, demonstrating the promise of this hybrid code generation paradigm. 
\end{itemize}

\paragraph{\textbf{Structure of the Paper}}
Section \ref{sec:approach} details the workflow of \iecoregen. 
Section \ref{sec:evaluation} outlines the experimental process and analyzes the results. 
Section \ref{sec:relatedwork} revisits existing studies on code generation. The final section provides conclusions and suggests directions for future research.

\section{Approach}\label{sec:approach}
\begin{figure}[!tb]
    \centering
    \includegraphics[width=1\linewidth]{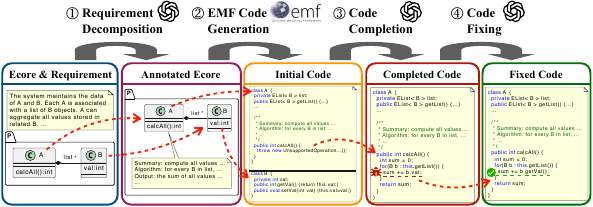}
    \caption{Workflow of \iecoregen}
    \label{fig:workflow}
\end{figure}

Figure \ref{fig:workflow} illustrates the overall process of \iecoregen. 
It starts with an Ecore model and a natural language requirement, breaking down this requirement into method specifications for the Ecore operations whose bodies are not specified. 
These specifications are annotated onto the model. 
The EMF generator then translates this annotated model into Java code, converting annotations into Java docstrings. 
\iecoregen subsequently uses these docstrings to guide LLMs in completing the unimplemented Java methods, checking for and correcting any compilation errors. 
We provide more details on each step below.

\subsection{Requirement Decomposition}
Existing LLM-based class and project-level code generation approaches \cite{classeval,javabench} generally require detailed class and method specifications.
In contrast, \iecoregen, derived from model-driven methodologies, utilizes an Ecore model alongside a natural language requirement as input.
An Ecore model is a simplified class diagram supported by EMF, which encompasses the modeling concepts of \textit{packages}, \textit{classes}, \textit{attributes}, \textit{references}, \textit{operations}, and \textit{inheritances}.

To facilitate the subsequent code generation, the first step of \iecoregen is \textit{requirement decomposition}.
This step aims to break down the given requirement and formulate a method specification for each Ecore operation defined in the Ecore model.
The process is elaborated below.
\begin{enumerate}
    \item Initially, we convert the Ecore model into PlantUML code, which is much easier for LLMs to understand.
    \item Following that, we provide LLMs with the PlantUML code and natural language requirements, instructing them to produce specifications for all operations. 
    We emphasize the importance of the \textit{modularization principle}, ensuring that each operation's functional boundary exhibits \textit{high cohesion and low coupling}. 
    Furthermore, LLMs are prompted to deduce any unspecified information, such as undefined constants, and are reminded to meticulously address every detail in the requirements.
    LLMs are requested to generate a structured method specification that covers the following aspects: 
    \begin{enumerate}
        \item \textbf{Summary} of the operation's functionality;
        \item \textbf{Algorithm} that realizes the required functionality;
        \item \textbf{Input} and \textbf{Output} that explain the meaning, format, and value ranges of the parameters and the return value of the operation;
        \item \textbf{Pre/Post-conditions} that must hold before/after the execution.
    \end{enumerate}
    
    \item Upon receiving responses from LLMs, we extract method specifications and store each as a model annotation of the corresponding operation in the Ecore model. Finally, we obtain an \textit{annotated Ecore model}.
\end{enumerate}

\subsection{EMF Code Generation}
The second step of \iecoregen is code generation using the EMF code generator.
We activate the EMF code generator to translate \textit{the annotated Ecore model} into \textit{initial Java code}.
The EMF code generator is created using Java Emitter Templates\footnote{JET Comprehensive Overview: \url{https://help.eclipse.org/latest/topic/org.eclipse.emf.doc/tutorials/jet/jet.html?cp=26\_1\_3}} (JET), a template language similar to Java Server Pages. 
It effectively translates the static features of an Ecore model, including class structures, attributes, and references, along with auxiliary functions such as getters, setters, object factories, and notifications, into Java code and docstrings. 
However, it typically treats all operations defined in the Ecore model as \textit{unimplementable}.
For each operation, the generator will generate the method signature and empty method bodies, as illustrated in Figure \ref{fig:codegen-mde}.
The model annotation for this operation will be printed as a docstring of the corresponding Java method.

Typically, EMF produces a Java interface and an implementation class for each Ecore class. 
\iecoregen simplifies this and asks the generator to produce only an implementation class by turning on the switch \textit{suppressInterfaces}.

\subsection{Code Completion}\label{sec:code-completion}
After the initial Java code is generated, \iecoregen performs the third step: \textit{code completion}.
Based on existing studies \cite{classeval,javabench}, we adopt the class-level holistic generation strategy, which processes Java classes one by one.

\begin{figure}[!tb]
    \centering
    \includegraphics[width=0.9\linewidth]{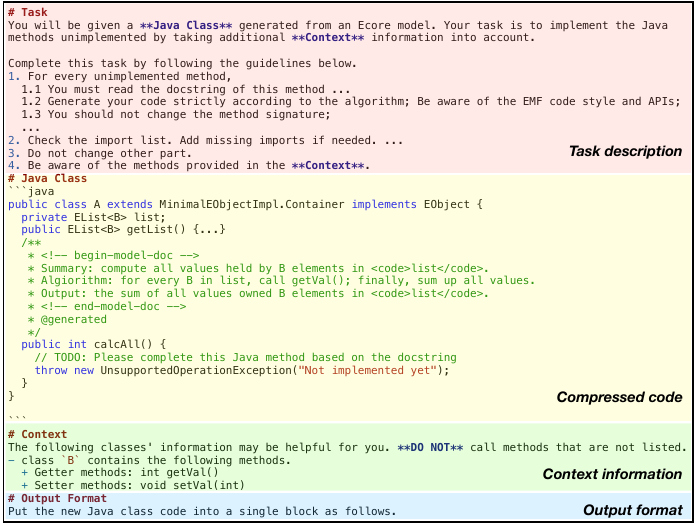}
    \caption{An example of code completion prompt}\label{fig:prompt:codecompletion}
\end{figure}

Our code completion process is \textit{guided} by the Ecore model.
We first traverse the Ecore model and collect the classes that own Ecore operations.
According to the class mapping rules, we locate the corresponding Java class files that need to be completed.
We load each class file and ask LLMs to complete the unimplemented Java methods, strictly following the docstring guidelines.
Figure \ref{fig:prompt:codecompletion} illustrates a prompt that is used to query LLMs.
LLMs can generate helper methods and include necessary imports but are restricted from modifying other parts of the class file. 
As LLMs-generated code must be compatible with the initial code, LLMs are instructed to adhere to EMF's coding style and APIs.
The completed code then replaces the initial version.

In this process, we must address the following technical issues. 
\begin{itemize}
    \item \textbf{Code compression}\quad    The initial Java class produced by EMF typically includes many docstrings, annotations, and code for reflection and notifications. Sending this initial code to LLMs can lead to rejection due to context length constraints. Therefore, we \textit{compress} the Java code before passing it to LLMs to reduce token usage. Initially, we use the Eclipse JDT parser to transform the code into an abstract syntax tree (AST). Next, we eliminate comment and docstring nodes, field initializers, and complete method bodies from the AST. Lastly, we serialize the modified AST back to Java and transmit this compact code to LLMs.
    
    \item \textbf{Context extraction}\quad  As these classes are interrelated, it is important to supply LLMs with details about related classes for accurate code completion. Without this \textit{context} information, LLMs might incorrectly refer to non-existent methods and APIs. Currently, \iecoregen employs a simple strategy for context extraction: it navigates the Ecore model to gather all Ecore classes related to the target class, such as its superclasses, associated classes, and data types for parameters and returns. The context is formed by joining the signatures of public methods from these related classes. We append this context to the initial code when querying LLMs, requesting that they take it into account.
    
    \item \textbf{Code replacement}\quad    Given that we send compressed code to LLMs for completion, directly substituting the initial code with the code generated by LLMs (which is compressed) is not feasible. Therefore, we perform a structured code merge as follows: First, we convert both the initial code and the LLM-generated code into ASTs for accurate code alignment. Next, we replace the methods intended for completion with the newly generated versions and append any new imports and helper methods. We keep the remaining parts unchanged to minimize unintended modifications.
\end{itemize}

\subsection{Code Fixing}
\begin{figure}[!tb]
    \centering
    \includegraphics[width=0.9\linewidth]{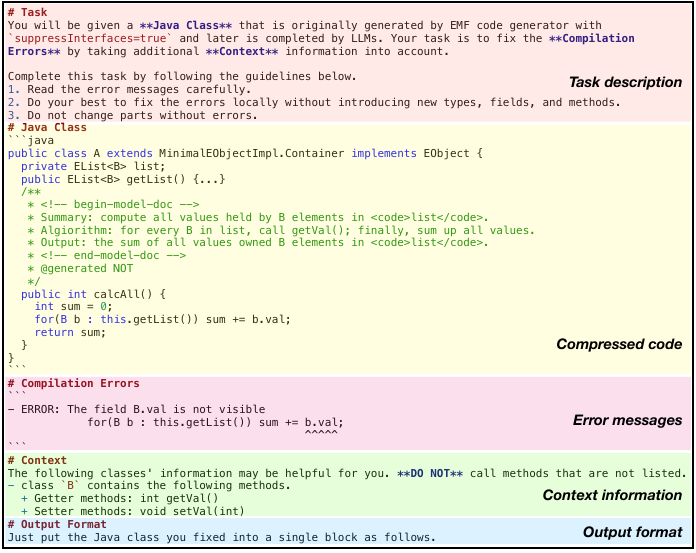}
    \caption{An example of code fix prompt}\label{fig:prompt:codefix}
\end{figure}

After we merge the LLM-generated code into the initial code, we proceed to compile the resulting code to ascertain the presence of any compilation errors.
Should errors be detected, we ask LLMs to fix them.
Figure \ref{fig:prompt:codefix} presents an example of an LLM prompt that outlines the errors found and requires LLMs to generate correct code.
This fixing process is iterated until either all compilation errors are resolved or the maximum number of retry attempts is reached.

To detect compilation errors, we employ the compiler provided by Eclipse JDT to compile the generated Java files.
We gather the compiler's output and parse out error messages, each consisting of the file path, the error type, the line number, the erroneous line of code, and the error explanation.

Subsequently, we utilize these error messages to pinpoint the Java classes and methods that exhibit compilation errors. 
The code compression procedure, as elaborated in Section \ref{sec:code-completion}, is performed again with a focus on these erroneous classes and methods. Additionally, we extract the contextual information and supply it to LLMs, as illustrated in Figure \ref{fig:prompt:codefix}.
Finally, the structured code merge is performed again to replace the erroneous code with the fixed version.

\section{Evaluation}\label{sec:evaluation}
The goal of this section is to evaluate \iecoregen with respect to the effectiveness of code generation concerning multiple classes from the perspective of academic researchers in the context of LLM-based software engineering.

We focus on the following three research questions.
\begin{itemize}
    \item \textbf{RQ1:} \textit{How does \iecoregen compare to pure LLM-based baselines in terms of the compilation and functional correctness of the generated code?}

    \begin{itemize}
        \item[] \textbf{Rationale:} Because LLM-based code generation is prone to hallucinations, comparing \iecoregen with LLM-only methods shows whether this hybrid approach---combining traditional model-driven techniques with modern LLMs---produces more correct multi-class code.
    \end{itemize}
    
    \item \textbf{RQ2:} \textit{How do different steps and components in \iecoregen contribute to the quality of generated code?}

    \begin{itemize}
        \item[] \textbf{Rationale:} 
        \iecoregen contains four key components: \textit{requirement decomposition}, \textit{code compression}, \textit{context extraction}, and \textit{code fixing}. 
        Omitting each component in turn lets us quantify its impact on performance.
    \end{itemize}

    \item \textbf{RQ3:} \textit{What are the common errors in the code generated by \iecoregen?}
  
    \begin{itemize}
        \item[] \textbf{Rationale:}
        Unlike approaches based purely on LLMs, \iecoregen combines MDE templates with neural generation, thereby introducing new failure modes. 
        An analysis and categorization of these errors expose limitations of \iecoregen and suggest avenues for improvement.
    \end{itemize}
\end{itemize}

\paragraph{\textbf{Tool and Data Availability}}
We have implemented a prototype tool based on EMF, Xtext, and Spring AI. Our tool and the raw data of our evaluation are available at \url{https://github.com/zzzzzzaplle/iEcoreGen}.

\subsection{Study Design}

\paragraph{\textbf{Metrics}} To assess the performance of code generation, we adopt the two most popular metrics, namely, \passk and \compk. 
In a code generation scenario, \passk is the expected value over problems of the probability that one or more of the $k$ generated code solutions will pass all tests \cite{passk}, while \compk is the expected probability that at least one of the $k$ randomly selected candidate solutions successfully compiles \cite{javabench}. 
They reflect compilation and functional correctness, respectively.
Both \passk and \compk are computable using Equation (\ref{eq:codemetrick}), where $n$ is the total number of solutions returned by LLMs, $k$ is the chosen number of solutions, and $c$ is the number of correct or compilable solutions. 
Referring to previous studies \cite{javabench}, we use $n=5$ and $k=1,3$.
\begin{equation}
\resizebox{0.55\textwidth}{!}{$
    \passk/\compk=\mathbb{E}_{problems}\left[1-\frac{\binom{n - c}{k}}{\binom{n}{k}}\right]   \label{eq:codemetrick}
$}
\end{equation}

\paragraph{\textbf{Benchmark}}
Due to the lack of well-defined and publicly available benchmarks for project-level code generation with software design models, we have constructed a benchmark for our evaluation as follows.
We first collected 130 domain descriptions from the literature \cite{mlset, Bozyigit2024} that were proposed for the evaluation of automated domain modeling.
We removed descriptions that were overly simple, unclear, or unrelated to software systems, selecting 20 suitable descriptions from different fields, with 6 from \cite{mlset} and 14 from \cite{Bozyigit2024}.

These descriptions were edited to fix typos, inconsistencies, and formatting errors. 
They focus on domain concepts and relationships and often lack detailed, computable functional requirements. 
Thus, we added between 3 to 6 functional requirements per description, totaling 20 system requirements. 
For each, we designed a class diagram (an Ecore model) incorporating domain concepts, key attributes, references, and operations reflecting the functional needs. 
Finally, we devised 5--6 natural language test cases for each functional requirement, amounting to 15--31 test cases per system.
Overall, our benchmark consists of 20 problems for code generation, each of which is equipped with a system requirement (containing 122--483 words and 3--6 functional items to be implemented), a class diagram in the Ecore format (containing 4--11 classes, 0--23 attributes, 2--11 references, and 4--12 operations), and 15--31 NL test cases. 

\paragraph{\textbf{Baselines}}
We compare \iecoregen with pure LLM-based baseline approaches, as summarized in Table \ref{tab:baselines}.
\textit{Base-R} generates source code from system requirements only.
In this approach, we send the requirement text to LLMs and ask them to generate \textit{a single Java source file} that implements the requirement.
\textit{Base-R+CD} generates the source file from both the requirements and the corresponding class diagram.
We convert an Ecore model into PlantUML before sending it to LLMs.
\textit{Base-R+CD+Fix} is similar to \textit{Base-R+CD} but is equipped with an additional code fixing step:
for the code generated by \textit{Base-R+CD}, we invoke a Java compiler and collect compilation errors;
afterward, we send the code along with the error messages to LLMs and ask them to fix the errors.
Because our approach is non-agentic, we do not include agent-based  baselines.
\begin{table}[!tb]
    \centering
    \caption{Baseline approaches}\label{tab:baselines}
    \resizebox{0.9\linewidth}{!}{
    \begin{tabular}{p{80pt}|p{300pt}}
    \toprule
        \textbf{Approach ID} & \textbf{Description}\\
    \midrule
       Base-R  & Code generation from NL requirements only\\
       Base-R+CD  & Code generation from NL requirements and class diagrams\\
       Base-R+CD+Fix  & Code generation from NL requirements and class diagrams with code fixing \\
    \bottomrule
    \end{tabular}
    }
\end{table}

\paragraph{\textbf{Studied LLMs}}
For our evaluation, we selected leading open-source LLMs, detailed in Table \ref{tab:llm}. 
DeepSeek-V3.2-Exp's 671B parameters handle complex tasks effectively. 
Qwen3-Coder-480B-A35B-Instruct is designed for programming. 
GPT-OSS-120B excels in language knowledge. 
Mistral-Small-3.2-24B-Instruct-2506 strikes a balance between size and efficiency. 
Llama-4-Maverick-17B-128E-Instruct is compact yet capable, offering insights into the size-performance trade-off.
\begin{table}[!b]
    \centering
    \caption{Studied LLMs}\label{tab:llm} 
    \resizebox{0.75\linewidth}{!}{
    \begin{tabular}{p{160pt}|p{60pt}|p{80pt}}
    \toprule
     \textbf{Name} & \textbf{Size} & \textbf{Release Time} \\
    \midrule
     DeepSeek-V3.2-Exp & 671B & September, 2025\\
     Qwen3-Coder-480B-A35B-Instruct & 480B & May, 2025\\
     GPT-OSS-120B & 120B & August, 2025\\
     Mistral-Small-3.2-24B-Instruct-2506 & 24B & June, 2025\\
     Llama-4-Maverick-17B-128E-Instruct  & 17B & April, 2025\\
    \bottomrule
    \end{tabular}
    }
\end{table}

\paragraph{\textbf{Processes}}
We conducted our evaluation based on the following processes.

For RQ1, we used \iecoregen and three baselines to generate functional code from the studied LLMs. We generated 5 samples per benchmark problem, then sent the code and NL test cases to LLMs for test code creation. We compiled and tested the functional code with the test code, and calculated \passk and \compk for each approach and LLM.
    
For RQ2, we used the best-performing LLM from RQ1 and ran another generation-and-testing round with \iecoregen for an ablation study. We systematically disabled key components—requirement decomposition, code compression, context extraction, and code fixing—then computed the two metrics to evaluate how omitting each component affected \iecoregen.

For RQ3, we reviewed the code produced by \iecoregen and recorded the residual errors. 
    We pinpointed the recurring error patterns of \iecoregen.

\subsection{RQ1: Overall Effectiveness}

Table \ref{tab:overall-effectiveness} presents the comparison between \iecoregen and baseline approaches.
The first column lists the LLMs we studied, and the second column shows the code generation approaches.
The remaining columns present the values of \passk and \compk (k=1,3) achieved by an approach on a certain LLM. 

\begin{table}[!tb]
    \centering
    \caption{Comparison of overall effectiveness among different approaches}\label{tab:overall-effectiveness}
    \resizebox{0.95\linewidth}{!}{
    \begin{tabular}{p{100pt}|p{80pt}|p{60pt}<{\centering}|p{60pt}<{\centering}|p{60pt}<{\centering}|p{60pt}<{\centering}}
    \toprule
        \multicolumn{1}{c|}{\multirow{2}{*}{\textbf{LLM}}} & \multicolumn{1}{c|}{\multirow{2}{*}{\textbf{Approach}}} & \multicolumn{2}{c|}{\textbf{\passk}} & \multicolumn{2}{c}{\textbf{\compk}} \\
        \cline{3-6}
        & & \textbf{k=1} & \textbf{k=3} & \textbf{k=1} & \textbf{k=3} \\
    \midrule
    \multirow{4}{100pt}{Deepseek-V3.2-Exp} & \iecoregen & \textbf{0.75} & \textbf{0.89} & \textbf{1} & \textbf{1} \\
         & \text{Base-R} & 0.47 \valdown{37\%} & 0.65 \valdown{27\%} & 0.88 \valdown{12\%} & 0.9 \valdown{10\%}\\
         & \text{Base-R+CD} & 0.54 \valdown{28\%} & 0.73 \valdown{18\%} & 0.98 \valdown{2\%} & \textbf{1} \valpar \\
         & \text{Base-R+CD+Fix} & 0.58 \valdown{23\%} & 0.75 \valdown{16\%} & \textbf{1} \valpar & \textbf{1} \valpar \\
    \midrule
    \multirow{4}{100pt}{Qwen3-Coder-480B-A35B-Instruct} & \iecoregen & \textbf{0.71} & \textbf{0.86} & 0.96 & \textbf{1} \\
         & \text{Base-R} & 0.5 \valdown{30\%} & 0.63 \valdown{27\%} & 0.97 \valup{1\%} & \textbf{1} \valpar \\
         & \text{Base-R+CD} & 0.57 \valdown{20\%} & 0.76 \valdown{12\%} & 0.91 \valdown{5\%} & \textbf{1} \valpar\\
         & \text{Base-R+CD+Fix} & 0.64 \valdown{10\%} & 0.76 \valdown{12\%} & \textbf{1} \valup{1\%} & \textbf{1} \valpar\\
    \midrule
    \multirow{4}{100pt}{GPT-OSS-120B} & \iecoregen & \textbf{0.7} & \textbf{0.86} & \textbf{0.99} & \textbf{1} \\
         & \text{Base-R} & 0.34 \valdown{51\%} & 0.58 \valdown{33\%} & 0.76 \valdown{23\%} & 0.94 \valdown{6\%} \\
         & \text{Base-R+CD} & 0.37 \valdown{47\%} & 0.62 \valdown{28\%} & 0.81 \valdown{18\%} & 0.93 \valdown{7\%} \\
         & \text{Base-R+CD+Fix} & 0.54 \valdown{23\%} & 0.69 \valdown{20\%} & {0.97} \valdown{2\%} & \textbf{1} \valpar \\
    \midrule
    \multirow{4}{100pt}{Mistral-Small-3.2-24B-Instruct-2506} & \iecoregen &  \textbf{0.50} & \textbf{0.73} & 0.89 & 0.95 \\
         & \text{Base-R} & 0.24 \valdown{52\%}  & 0.47 \valdown{36\%} & 0.81 \valdown{9\%} & 0.99 \valup{4\%}\\
         & \text{Base-R+CD} & 0.33 \valdown{34\%} & 0.49 \valdown{33} & 0.92 \valup{3\%} & \textbf{1} \valup{5\%}\\
         & \text{Base-R+CD+Fix} & 0.39 \valdown{22\%} & 0.61 \valdown{16\%} & \textbf{0.99} \valup{11\%} & \textbf{1} \valup{5\%}\\
    \midrule
    \multirow{4}{100pt}{Llama-4-Maverick-17B-128E-Instruct} & \iecoregen & \textbf{0.57} & \textbf{0.83} & 0.93 & 0.98 \\
         & \text{Base-R} & 0.35 \valdown{39\%} & 0.53 \valdown{36\%} & 0.96 \valup{3\%} & \textbf{1} \valup{2\%}\\
         & \text{Base-R+CD} & 0.45 \valdown{21\%} & 0.74 \valdown{11\%} & {0.96} \valup{3\%} & \textbf{1} \valup{2\%}\\
         & \text{Base-R+CD+Fix} & 0.54 \valdown{5\%} & 0.74 \valdown{11\%} & \textbf{0.98} \valup{5\%} & \textbf{1} \valup{2\%} \\
    \bottomrule
    \end{tabular}
    }
\end{table}

It is evident that \iecoregen consistently achieves higher \passk scores than all baseline methods across every evaluated LLM, as shown in Table \ref{tab:overall-effectiveness}.
Overall, \iecoregen yields improvements of roughly 5\%--52\% in pass@1 (29\% on average) and 11\%--36\% in pass@3 (22\% on average), respectively.
This suggests that \iecoregen enhances the likelihood of producing functionally correct code by integrating template-based code generation with LLM-based methods.
We attribute this improvement to the core principle of \iecoregen: by offloading the generation of a foundational, correct initial Java code to code templates in MDE, the task for LLMs is reduced to completing critical missing parts, thereby constraining their problem-solving scope and mitigating hallucination risks.

Regarding \compk, \iecoregen is generally comparable with the baselines.
For larger language models, such as Deepseek-V3.2-Exp, Qwen3-Coder-480B-A35B-Instruct, and GPT-OSS-120B, the \compk performance of \iecoregen is comparable to that of the baselines. 
However, for smaller models, including Mistral-Small-3.2-24B-Instruct-2506 and Llama-4-Maverick-17B-128E-Instruct, \iecoregen yields slightly lower \compk scores than the baselines.
This degradation stems from the stricter syntactic constraints that are imposed on code generation by \iecoregen.
LLMs must respect EMF-specific coding conventions to produce syntactically valid code.
For instance, in EMF, the getter of a boolean attribute ``X'' must be named ``isX()'' rather than ``getX()''.
Smaller LLMs, in particular, may lack sufficient EMF-specific knowledge due to the limited presence of EMF-related code in their training data.
As a result, in comparison to baselines that place fewer restrictions on code generation, \iecoregen can slightly increase the compilation error rate.
Currently, we explicitly instruct LLMs to follow EMF's coding style and APIs, as illustrated in Figure \ref{fig:prompt:codecompletion}.
Future work could focus on how to provide MDE knowledge for LLMs.

Table \ref{tab:overall-effectiveness} shows that, for all generation strategies, Deepseek and Qwen3-Coder consistently rank among the top-performing LLMs, whereas Mistral-Small-3.2 is the weakest model.
Furthermore, \iecoregen yields strong results on Deepseek, Qwen3-Coder, and GPT-OSS, but performs poorly on Mistral and Llama-4.


\paragraph{\textbf{Answer to RQ1}}
From Table \ref{tab:overall-effectiveness}, we conclude for RQ1 that \textbf{\iecoregen improves the functional correctness of code generation} while preserving similar compilation correctness.

\subsection{RQ2: Ablation Study}

To investigate the contributions of key components in \iecoregen (i.e., \textit{requirement decomposition}, \textit{code compression}, \textit{context extraction}, and \textit{code fixing}), we conducted an ablation study using Deepseek-V3.2-Exp, the language model on which \iecoregen achieved its best performance in the previous experiments. The results are summarized in Table \ref{tab:ablation}. 

\begin{table}[!b]
    \centering
    \caption{Results of ablation experiment on Deepseek-V3.2-Exp}\label{tab:ablation}
    \resizebox{0.95\linewidth}{!}{
    \begin{tabular}{p{160pt}|p{60pt}<{\centering}|p{60pt}<{\centering}|p{60pt}<{\centering}|p{60pt}<{\centering}}
    \toprule
        \multicolumn{1}{c|}{\multirow{2}{*}{\textbf{Approach}}} & \multicolumn{2}{c|}{\textbf{\passk}} & \multicolumn{2}{c}{\textbf{\compk}} \\
        \cline{2-5}
         & \textbf{k=1} & \textbf{k=3} & \textbf{k=1} & \textbf{k=3} \\
    \midrule
        \iecoregen & {0.75} & {0.89} & 1 & {1}\\
    \midrule
        \quad --- w/o requirement decomposition & 0.32\valdown{57\%} & 0.52\valdown{41\%} & 0.77\valdown{23\%} & 0.92\valdown{8\%} \\
        \quad --- w/o code compression          & 0.47\valdown{37\%} & 0.69\valdown{22\%} & 0.93\valdown{7\%} & 0.98\valdown{2\%} \\
        \quad --- w/o context extraction        & 0.30\valdown{60\%} & 0.54\valdown{39\%} & 0.55\valdown{45\%} & 0.72\valdown{28\%} \\
        \quad --- w/o code fixing               & 0.38\valdown{53\%} & 0.53\valdown{40\%} & 0.65\valdown{35\%} & 0.81\valdown{19\%}\\
    \bottomrule
    \end{tabular}
    }
\end{table}

The results clearly show that removing any major component substantially degrades performance on all metrics.  
(1) Dropping requirement decomposition severely reduces functional correctness, with pass@1 and pass@3 falling by 57\% and 41\%, indicating that decomposing complex tasks into smaller steps is essential for steering the LLM toward correct solutions.  
(2) Turning off code compression lowers pass@1 and pass@3 by 37\% and 22\%, suggesting that concise code improves LLM reasoning. EMF-generated code typically includes numerous comments and auxiliary EMF-specific code; without compression, this extra material can distract LLMs from completion and repair tasks, increasing errors.  
(3) Removing context extraction has the most severe impact, cutting pass@1 by 60\% and compilation@1 by 45\%. Supplying surrounding class context, including the information on related classes, helps LLMs avoid hallucinations, particularly the incorrect method calls.  
(4) Skipping code fixing markedly hurts the overall performance---\passk falls by 53\%/40\% and \compk by 35\%/19\% (k=1,3)---showing that this post-processing stage is vital for correcting the LLM’s subtle initial mistakes.

\paragraph{\textbf{Answer to RQ2}}
This ablation study addresses RQ2 by showing that all four core components of \iecoregen are individually crucial and complementary, and that only their joint use yields the \iecoregen's best performance.


\subsection{RQ3: Error Cases Exploration}
Although \iecoregen yields higher \passk scores, it can still generate code that fails to compile or functions incorrectly.
To further analyze the typical failure modes of \iecoregen, we conducted a manual inspection of the code produced by Deepseek-V3.2-Exp and Qwen3-Coder-480B-A35B-Instruct.

In total, we reviewed 200 code samples, including 4 non-compilable samples, 50 error samples that failed at runtime, and 145 unique error cases.
We analyzed and classified every error case that emerged during compilation or testing.
Table \ref{tab:case-classification} shows the classification of these cases.
The most frequent compilation error is \textit{unresolved symbols}, where the generated code references undefined fields, methods, or enumeration literals. The most frequent testing error is \textit{requirement omission}, where some required functions are missing from the code.
\begin{table}[!b]
    \centering
    \caption{Classification of error cases}\label{tab:case-classification}
    \resizebox{1\linewidth}{!}{
    \begin{tabular}{p{50pt}|p{100pt}|p{240pt}|p{20pt}p{40pt}}
    \toprule
        \textbf{Phase} & \textbf{Category} & \textbf{Description} & \multicolumn{2}{l}{\textbf{Count}}\\
    \midrule
        \multirow{2}{*}{Compilation}
            & Unresolved Symbols    & Use undefined fields, methods, or enumeration literals & 4 & (2.8\%)\\
            & Type Incompatible     & The types of left and right values are incompatible & 1 & (0.7\%)\\
    \midrule
        \multirow{9}{*}{Testing}    
            & Requirement Omission  & A certain requirement is omitted                      & 62 & (42.8\%)\\
            & Operation Misuse      & A wrong operation/API is invoked or wrongly invoked   & 21 & (14.5\%)\\
            & Distorted Requirements& Code is generated based on distorted requirements     & 18 & (12.4\%)\\
            & Boundary Case Handling& Fail to handle edge/boundary cases correctly          & 18 & (12.4\%)\\ 
            & Inverted Condition    & A branch condition is wrongly inverted                & 6 & (4.1\%)\\
            & Fake Requirements     & The generated logic is not required or specified      & 5 & (3.4\%)\\
            & Constraint Violation  & Some business constraints, such as ordering, are neglected & 5 & (3.4\%)\\
            & Others                & Other runtime errors                                  & 5 & (3.4\%)\\ 
    \bottomrule
    \end{tabular}}
\end{table}

\begin{figure}[!tb]
    \subfloat[Requirement omission\label{fig:error:reqomit}]{\includegraphics[width=\textwidth]{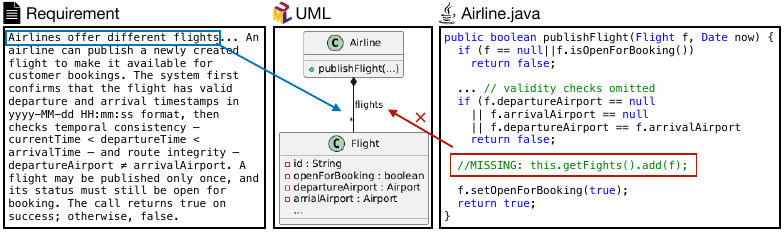}}

    \subfloat[Operation misuse\label{fig:error:opmiss}]{\includegraphics[width=\textwidth]{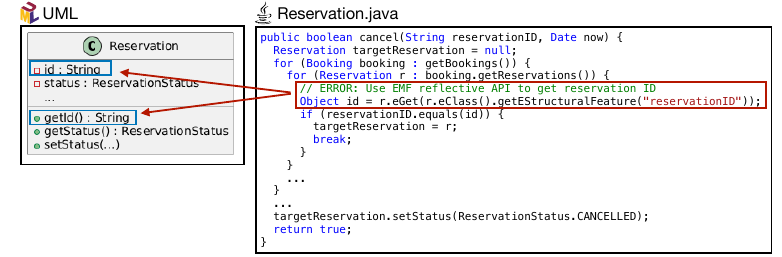}}

    \caption{Error case examples}\label{fig:error-cases}
\end{figure}

Figure \ref{fig:error:reqomit} illustrates a specific case of \textit{requirement omission}.
The requirements specify that ``\textit{Airlines offer different flights}'' and ``\textit{An airline can publish a flight}''.
Therefore, in \texttt{publishFlight(Flight f, Date now)}, the flight \texttt{f} should be inserted into the airline’s \texttt{flights} collection.
However, the generated code fails to do this, suggesting that the requirement decomposition step is improvable. 

Figure \ref{fig:error:opmiss} shows an example of \textit{operation misuse}.
The code invokes the reflective APIs of EMF to retrieve the \texttt{id} of a reservation object.
Nevertheless, it should either call \texttt{getId()} directly or use \texttt{"id"} (rather than \texttt{"reservationID"}) in the reflective API call, i.e., \texttt{r.eClass().getEStructuralFeature("id")}.
This highlights LLMs’ shallow understanding of EMF APIs and coding style.

\paragraph{\textbf{Answer to RQ3}} Table \ref{tab:case-classification} shows that \textit{requirement omission} and \textit{operation misuse} are the most common errors, highlighting the room for further improvement.

\subsection{Threats to Validity}
\noindent\textbf{- Accuracy of test code} The test code we used to verify the generated code is still produced by LLMs, which may also be incorrect. To mitigate this issue, we made every effort to review each failed test case to ensure its correctness.

\noindent\textbf{- Benchmark complexity} 
The benchmark problems are fairly simple, with requirements capped at 483 words and at most 11 classes, so they may not represent real-world projects well. 
Yet, the baseline pass@1 score is only 0.46 on average, which is unsatisfactory. 
We will evaluate more realistic problems.

\noindent\textbf{- Omission of closed-source LLMs} We evaluated only open-source LLMs because of time and cost constraints. Although closed-source models should ideally be included, the selected open-source LLMs perform comparably on many SE tasks, so we expect our findings to generalize to closed-source LLMs.

\section{Related Work}\label{sec:relatedwork}

In model-driven engineering (MDE), traditional code generation typically depends on precisely specified transformation rules and code templates \cite{mdecg,cgbymt}, which traverse input models in a systematic manner and translate them into source code. 
This approach underlies numerous MDE frameworks, including EMF \cite{emf}, Papyrus \cite{papyrus}, MontiThings \cite{MontiThings}, MontiAnna \cite{MontiAnna}, and ML-Quadrat \cite{Moin_2022}.
Yet, it rapidly reaches its limits when the desired functionality is not covered by the predefined templates.
This work aims to broaden the scope of conventional approaches through the integration of LLMs.




The LLM-based coding paradigm has attracted substantial research attention in recent years. 
These generative neural networks, such as Deepseek \cite{deepseekai2024deepseekv3technicalreport}, have demonstrated remarkable capabilities in many code-related tasks, including code generation \cite{cg-Qingyao,cg-Xiangyu,cg-Xiaoqing, Zhang2025}, test generation \cite{tc-Jain2025,tc-Navarro2026,tc-Xu2025,tc-Yang2025}, and bug fixing and repair \cite{bf-Mavalankar2025,bf-Peng2024,bf-Xunzhu,bf-Yuze}.
Existing LLM-based code generation methods typically take NL requirements or specifications as input \cite{wei2024requirements, bf-Ding2024,flcg-Wang2024,flcg-Wu2024} and use LLMs to iteratively translate these NL descriptions into code.
\iecoregen, introduced in this work, uses templates to construct code skeletons and delegates only the unfinished methods to LLMs. By narrowing the subproblems LLMs must address, it enhances their effectiveness in generating complex, multi-class code. 

LLMs are also applied to address model-related challenges \cite{DiRocco2025}, such as model generation \cite{chen23,herwanto2024automating,apvrille2024system,camara2023assessment,chen2023use,10.1007/978-3-031-94569-4_1-modeling-quality-Lris} and model correctness checks \cite{ahmed2025mcetbehavioralmodelcorrectness}.
Boubou et al. \cite{Niang2025} leveraged model-driven reverse engineering to enhance LLM-based code generation. 
Their approach extracts abstract models containing core components and dependencies from existing codebases and then uses structured information, such as method signatures, to construct detailed prompts that guide LLMs in generating code that aligns with project structure and requirements.
To the best of our knowledge, \textbf{we are among the first to introduce a hybrid method that integrates traditional template-based code generation with LLM-based generation} to more effectively combine LLMs with MDE.

\section{Conclusion}
This paper explores a novel and hybrid code generation approach, \iecoregen, that integrates conventional template-based technologies in MDE and LLMs.
By employing code templates to construct correct initial code and using LLMs to address complex and flexible functional requirements, \iecoregen integrates the strengths of both paradigms. 
Empirical results indicate that \iecoregen surpasses pure-LLM baselines on \passk while maintaining comparable performance on \compk.

In the future, we plan to enhance \textit{requirement decomposition} and \textit{context extraction} of \iecoregen, as the error analysis has highlighted clear opportunities for improvement.
We will further strengthen our evaluation by incorporating closed-source LLMs and more complicated benchmarks.
Finally, we will improve the tool support and submit it as an extension to EMF.

\begin{credits}

\subsubsection{\discintname}
The authors have no competing interests to declare that are relevant to the content of this article.
\end{credits}
%
%
%
\bibliographystyle{splncs04}
\bibliography{reference.bib}

\end{document}